\documentstyle[12pt]{article}
\begin{document}

\title{Shear-free radiating collapse and conformal flatness}

\author{ L. Herrera$^1$\thanks{e-mail: laherrera@telcel.net.ve}, G. Le
Denmat$^2$\thanks{e-mail: gele@ccr.jussieu.fr}, N. O.
Santos$^{2,3,4}$\thanks
{e-mail: nos@cbpf.br} and A. Wang$^{5,6}$\thanks{e-mail:
aw@guava.physics.uiuc.edu}\\
\small{$^1$Escuela de F\'{\i}sica, Facultad de Ciencias,} \\
\small{Universidad Central de Venezuela, Caracas, Venezuela.}\\
\small{$^2$LERMA/CNRS-FRE 2460, Universit\'e Pierre et Marie Curie, ERGA,}\\
\small{Bo\^{\i}te 142, 4 Place Jussieu, 75005 Paris Cedex 05, France.}\\
\small{$^3$Laborat\'orio Nacional de Computa\c{c}\~ao Cient\'{\i}fica,
25651-070 Petr\'opolis RJ, Brazil.}\\
\small{$^4$Centro Brasileiro de Pesquisas F\'{\i}sicas, 22290-180 Rio de
Janeiro RJ, Brazil.}\\
\small{$^5$Departamento de F\'{\i}sica Te\'orica, Universidade do Estado do
Rio de Janeiro,}\\
\small{20550-013 Rio de Janeiro RJ, Brazil.}\\
\small{$^6$Department of Physics, University of Illinois at Urbana-Champaign,}\\
\small{1110 West Green Street, Urbana, IL 61801-3080, United States.}}
\maketitle

\begin{abstract}
Here we study some general properties of spherical shear-free collapse. Its
general solution when imposing conformal flatness is reobtained
\cite{Som,Maiti} and matched  to the
outgoing Vaidya spacetime. We propose a simple model satisfying these
conditions and study its physical consequences. Special attention deserve,
the role played by relaxational processes and the conspicuous link betweeen
dissipation and density
inhomogeneity.
\end{abstract}

\maketitle

\newpage

\section{Introduction}
Gravitational collapse of stars is an important problem of astrophysics and
building realistic models of collapse remains a formidable task. One
interesting problem is to add heat
 flow to spherically symmetric models.

 Indeed, dissipation due to the emission of massless
particles (photons and/or neutrinos) is a characteristic process in the
evolution of massive stars. In fact, it seems that the only plausible
mechanism to carry away the bulk of the binding energy of the collapsing
star, leading to a neutron star or black hole is neutrino emission
\cite{1}.

In the diffusion approximation, it is assumed that the energy flux of
radiation (as that of
thermal conduction) is proportional to the gradient of temperature. This
assumption is in general very sensible, since the mean free path of
particles responsible for the propagation of energy in stellar
interiors is in general very small as compared with the typical
length of the object.
Thus, for a main sequence star as the sun, the mean free path of
photons at the centre, is of the order of $2\, cm$. Also, the
mean free path of trapped neutrinos in compact cores of densities
about $10^{12} \, g.cm.^{-3}$ becomes smaller than the size of the stellar
core \cite{3,4}.

Furthermore, the observational data collected from supernovae 1987A
indicates that the regime of radiation transport prevailing during the
emission process, is closer to the diffusion approximation than to the
streaming out limit \cite{5}.

 Many solutions of Einstein's field
equations with dissipative fluids carrying heat flow have been studied (see
\cite{Bonnor} for references
up to
1989 and \cite{Schafer,HS, chan} for more recent ones).

In this vein here we study dissipative spherical collapse with shear-free
motion. Spherical conformally flat fluids undergoing dissipation in the
form of radial heat flow where first
considered in \cite{Som} and generalized in \cite{Maiti}. Here we reobtain
the general conformally flat solution in a slightly different way. We match
this spacetime to a radiating
null field described by the outgoing Vaidya spacetime. A simple model is
considered satisfying these conditions.

The paper is organized as follows. In section 2 the field equations are
presented; in section 3 we reobtain the general solution by considering
conformal flatness of spacetime; in
section
 4 we state the junction conditions to the external outgoing Vaidya null
radiating field; section 5 presents a simple collapsing dissipative model
and we finish with a brief conclusion.

\section{Field equations}
We assume a sphere of collapsing perfect fluid with heat flow. Its spherical
surface $\Sigma$ has center 0 and is filled with radially moving perfect
fluid conducting heat flow, so having energy momentum tensor
\begin{equation}
T_{\alpha\beta}=\left(\mu+p\right)w_{\alpha}w_{\beta}+pg_{\alpha\beta}+
q_{\alpha}w_{\beta}+w_{\alpha}q_{\beta}, \label{I1}
\end{equation}
where $\mu$ and $p$ are the proper density and pressure of the fluid,
$w_{\alpha}$ its unit four-velocity, $q_{\alpha}$ the heat conduction
satisfying $q_{\alpha}w^{\alpha}=0$ and $g_{\alpha\beta}$ is the metric
tensor of spacetime.

We choose comoving coordinates within $\Sigma$ and impose shear-free fluid
motion
which allows the metric be written in the form \cite{Glass}
\begin{equation}
ds^2=-A^2dt^2+B^2\left[dr^2+r^2\left(d\theta^2+\sin^2\theta
d\phi^2\right)\right],
\label{I2}
\end{equation}
where $A$ and $B$ are only functions of $r$ and $t$. We number the
coordinates $x^0=t$, $x^1=r$, $x^2=\theta$ and $x^3=\phi$ and then we
have the four-velocity given by
\begin{equation}
w_{\alpha}=-A\delta_{\alpha}^0, \label{I3}
\end{equation}
and the heat flows radially,
\begin{equation}
q^{\alpha}=q\delta_1^{\alpha}, \label{I4}
\end{equation}
where $q$ is a function of $r$ and $t$.

The rate of collapse $\Theta={w^{\alpha}}_{;\alpha}$ of the fluid sphere
is given, from (\ref{I2}) and (\ref{I3}), by
\begin{equation}
\Theta=3\frac{\dot{B}}{AB}, \label{I5a}
\end{equation}
where the dot stands for differentiation with respect to $t$.

The spacetime described by (\ref{I2}) has the following non-null components
of the Weyl tensor $C_{\alpha\beta\gamma\delta}$,
\begin{equation}
C_{2323}=\frac{r^4}{3}B^2\sin^2\theta\left[\left(\frac{A^{\prime}}{A}-
\frac{B^{\prime}}{B}\right)\left(\frac{1}{r}+2\frac{B^{\prime}}{B}\right)-
\left(\frac{A^{\prime\prime}}{A}-\frac{B^{\prime\prime}}{B}\right)\right],
\label{I6}
\end{equation}
and
\begin{eqnarray}
C_{2323}=-r^4\left(\frac{B}{A}\right)^2\sin^2\theta C_{0101}
=2r^2\left(\frac{B}{A}\right)^2\sin^2\theta C_{0202} \nonumber \\
=2r^2\left(\frac{B}{A}\right)^2C_{0303}=-2r^2\sin^2\theta C_{1212}=-2r^2C_{1313},
\label{I6a}
\end{eqnarray}
where the primes stand for differentiation with respect to $r$.

The non null components of Einstein's field equations
$G_{\alpha\beta}=\kappa T_{\alpha\beta}$, where $G_{\alpha\beta}$ is the
Einstein tensor and $T_{\alpha\beta}$ is given by (\ref{I1}), with metric
(\ref{I2}) are
\begin{eqnarray}
G_{00}=-\frac{A^2}{B^2}\left[2\frac{B^{\prime\prime}}{B}-
\left(\frac{B^{\prime}}{B}\right)^2+\frac{4}{r}\frac{B^{\prime}}{B}\right]+
3\left(\frac{\dot{B}}{B}\right)^2=\kappa\mu A^2, \label{I7} \\
G_{11}=\left(\frac{B^{\prime}}{B}\right)^2+\frac{2}{r}\frac{B^{\prime}}{B}+
2\frac{A^{\prime}}{A}\frac{B^{\prime}}{B}+\frac{2}{r}\frac{A^{\prime}}{A}
\nonumber \\
+\frac{B^2}{A^2}\left[-2\frac{\ddot{B}}{B}-\left(\frac{\dot{B}}{B}\right)^2+
2\frac{\dot{A}}{A}\frac{\dot{B}}{B}\right]=\kappa p B^2, \label{I8} \\
G_{22}=\frac{G_{33}}{\sin^2\theta}=r^2\left[\frac{A^{\prime\prime}}{A}+
\frac{1}{r}\frac{A^{\prime}}{A}+\frac{B^{\prime\prime}}{B}-
\left(\frac{B^{\prime}}{B}\right)^2+\frac{1}{r}\frac{B^{\prime}}{B}\right]
\nonumber \\
+r^2\frac{B^2}{A^2}\left[-2\frac{\ddot{B}}{B}-
\left(\frac{\dot{B}}{B}\right)^2
+2\frac{\dot{A}}{A}\frac{\dot{B}}{B}\right]=\kappa p r^2 B^2, \label{I9} \\
G_{01}=-2\left(\frac{\dot{B}}{AB}\right)^{\prime}A=-\kappa q A B^2. \label{I10}
\end{eqnarray}
>From (\ref{I10}) with (\ref{I5a})
we obtain
\begin{equation}
\kappa q B^2=\frac{2}{3}\Theta^{\prime}, \label{I10a}
\end{equation}
which shows that the outflow of heat, $q>0$, imposes $\Theta^{\prime}>0$,
meaning that, if $\Theta<0$, dissipation diminishes the rate of collapse
towards the outer layers of matter.
If $q=0$ then from (\ref{I10a})
$\Theta^{\prime}=0$ which means that collapse is homogeneous \cite{HS}.

The mass function $m(r,t)$ of Cahill and McVittie \cite{Cahill} is obtained
from the Riemann tensor component ${R_{23}}^{23}$ and it is
for metric (\ref{I2})
\begin{equation}
m(r,t)=\frac{\left(rB\right)^3}{2}{R_{23}}^{23}=\frac{r^3B}{2}\left[
\left(\frac{\dot{B}}{A}\right)^2-
\left(\frac{B^{\prime}}{B}\right)^2\right]-r^2B^{\prime}. \label{I11}
\end{equation}
Differentiating $m(r,t)$ with respect to $r$ and $t$ and considering the field
equations (\ref{I7}-\ref{I10}) we obtain
\begin{eqnarray}
m^{\prime}=\frac{\kappa}{2}\left[\mu (rB)^2 (rB)^{\prime}+
qr^3B^4\frac{\dot{B}}{A}\right], \label{I11a} \\
\dot{m}=-\frac{\kappa}{2}\left[pr^3B^2\dot{B}
+q(rB)^2(rB)^{\prime}A\right]. \label{I11b}
\end{eqnarray}
>From (\ref{I11a}) and (\ref{I11b}) we have that the heat flow diminishes
the gradient and the time derivative of $m(r,t)$. This agrees with the
discussion concerning (\ref{I10a}),
since dissipation diminishes the total amount of matter, it is expected
that the rate of collapse slows down. Furthermore, this agrees too with the
results obtained for the dynamical
instability of nonadiabatical spherical collapse \cite{Herrera,Bonnor}
where it is proved that relativistically dissipation diminishes instability
due to the decrease of matter content
inside a collapsing sphere.

The Riemann curvature tensor can be split into the Weyl tensor and parts
which involve only the Ricci tensor and the curvature scalar. This allows
to say that the Weyl part is
constructed
only by the gravitational field.
Considering the scalar of the Weyl tensor
\begin{equation}
{\mathcal C}^2=C_{\alpha\beta\gamma\delta}C^{\alpha\beta\gamma\delta},
\label{I12}
\end{equation}
with (\ref{I6}), (\ref{I7}) and (\ref{I11}) we obtain after a long
calculation
\begin{equation}
{\mathcal C}^2=48\left[\frac{m}{\left(rB\right)^3}-
\frac{\kappa\mu}{6}\right]^2=48\frac{m_{\mathcal C}^2}{(rB)^6}, \label{I13}
\end{equation}
where $m_{\mathcal C}$ is defined as the pure gravitational mass,
\begin{equation}
m_{\mathcal C}=m-\frac{\kappa}{6}\mu (rB)^3.
\label{I14}
\end{equation}

\section{Conformally flat solution}
Here we impose conformal flatness to the spacetime given by (\ref{I2}),
i.e. all its Weyl tensor components must be zero valued. From (\ref{I6}) and
(\ref{I6a}) we see that if $C_{2323}=0$ this condition is fulfilled, hence
we have
\begin{equation}
r\left(\frac{A^{\prime\prime}}{A}-\frac{B^{\prime\prime}}{B}\right)-
\left(\frac{A^{\prime}}{A}-\frac{B^{\prime}}{B}\right)
\left(1+2r\frac{B^{\prime}}{B}\right)=0. \label{II1}
\end{equation}
We can integrate (\ref{II1}) and after reparametrizing $t$ we obtain
\begin{equation}
A=\left[C_1\left(t\right)r^2+1\right]B, \label{II2}
\end{equation}
where $C_1$ is an arbitrary function of $t$.
>From the isotropy of pressure, (\ref{I8}) and (\ref{I9}),
equating $r^{-2}G_{22}-G_{11}$ to zero and using
(\ref{II2}) we find
\begin{equation}
\frac{B^{\prime\prime}}{B^{\prime}}-2\frac{B^{\prime}}{B}-\frac{1}{r}=0,
\label{II3}
\end{equation}
which is easily integrated,
\begin{equation}
B=\frac{1}{C_2(t)r^2+C_3(t)}, \label{II4}
\end{equation}
where $C_2$ and $C_3$ are arbitrary functions of $t$. The solution found
in \cite{Som} is a particular case of (\ref{II2}) and (\ref{II4}) with $C_1=0$.
All conformally flat perfect fluid solutions with $q=0$ have been obtained
by Stephani \cite{Stephani, Kramer}

Conformal flatness imposes ${\mathcal C}=0$ and from (\ref{I13}) we have
that the pure gravitational mass $m_{\mathcal C}=0$ and
\begin{equation}
m=\frac{\kappa}{6}\mu\left(rB\right)^3, \label{II5}
\end{equation}
which is similar to the result obtained in \cite{Lake} with $q=0$.
>From (\ref{I11a}) with (\ref{II5}) we have
\begin{equation}
\mu^{\prime}=qB^2\Theta, \label{II5a}
\end{equation}
which shows that for $q>0$ and $\Theta<0$ then $\mu^{\prime}<0$ implying that
the density diminishes with increasing $r$.
While from (\ref{I5a}) with (\ref{I10a}) we obtain
\begin{equation}
\kappa\mu^{\prime}=\frac{1}{3}(\Theta^2)^{\prime}, \label{I5b}
\end{equation}
which can be integrated, giving
\begin{equation}
\kappa\mu=\frac{\Theta^2}{3}+g(t), \label{I5c}
\end{equation}
where $g$ is a function only of $t$.

Substituting solution (\ref{II2}) and (\ref{II4}) into (\ref{I8}),
(\ref{I9}) and (\ref{I11}) we obtain,
\begin{eqnarray}
\kappa\mu=3\left(\frac{\dot{C}_2r^2+\dot{C}_3}{C_1r^2+1}\right)^2+12C_2C_3,
\label{II6} \\
\kappa p=\frac{1}{(C_1r^2+1)^2}\left[2(\ddot{C}_2r^2+\ddot{C}_3)(C_2r^2+C_3)-
3(\dot{C}_2r^2+\dot{C}_3)^2 \right. \nonumber \\
\left. -2\frac{\dot{C}_1}{C_1r^2+1}
(\dot{C}_2r^2+\dot{C}_3)(C_2r^2+C_3)r^2\right] \nonumber \\
+\frac{4}{C_1r^2+1}\left[C_2(C_2-2C_1C_3)r^2+C_3(C_1C_3-2C_2)\right],
\label{II7}\\
\kappa q=4(\dot{C}_3C_1-\dot{C}_2)
\left(\frac{C_2r^2+C_3}{C_1r^2+1}\right)^2r. \label{II8}
\end{eqnarray}
The expansion $\Theta$ of the fluid sphere given by (\ref{I5a}) with
(\ref{II2}), (\ref{II4}) and (\ref{II8}), is
\begin{equation}
\Theta=-3\frac{\dot{C}_2r^2+\dot{C}_3}{C_1r^2+1}=-3\left[\dot{C}_3-
\frac{\kappa q r}{4}\frac{C_1r^2+1}{(C_2r^2+C_3)^2}\right]. \label{II9}
\end{equation}
We see from (\ref{II9}) that if $q=0$ the contraction is homogeneous,
however if $q\neq 0$, dissipation produces inhomogeneous collapse, which
has already been remarked in (\ref{I10a}).

The density $\mu$ in (\ref{II6}) confirms the result (\ref{I5c}) with
$g(t)=12C_2C_3$.

It is possible to prove, after a long calculation, that the fluid
(\ref{II6}-\ref{II8}) does not satisfy an equation of state
of the form $p=c\mu$, where $c$ is a constant, with $q\neq 0$. A family of
solutions with heat flux satisfying an equation of state is given in
\cite{Wagh}.

In the next section we consider the junction conditions of the collapsing
dissipative fluid to a radiating field.

\section{Junction conditions}
If the collapsing fluid lies within a spherical surface $\Sigma$ it must be
matched to a suitable exterior. Since heat will be leaving the fluid across
$\Sigma$ the exterior is not vacuum, but the outgoing Vaidya spacetime which
models the radiation and has metric
\begin{equation}
ds^2=-\left[1-\frac{2m(v)}{\rho}\right]dv^2-2dvd\rho+\rho^2(d\theta^2+\sin^2
\theta
d\phi^2),
\label{III1}
\end{equation}
where $m(v)$ is the total mass inside $\Sigma$ and is a function of the retarded
time $v$. In (\ref{III1}) $\rho$ is a radial coordinate given in a
non-comoving frame.
The matching of these two spacetimes (\ref{I2}) and (\ref{III1}), using the
field
equations (\ref{I8}-\ref{I10}) and the mass function (\ref{I11}) satisfies
\cite{Santos,Bonnor}
\begin{eqnarray}
(rB)_{\Sigma}=\rho_{\Sigma}, \label{III2}\\
p_{\Sigma}=(qB)_\Sigma, \label{III3}\\
m(v)=\left\{\frac{r^3}{2}\left[\frac{\dot{B}^2B}{A^2}-\frac{(B^{\prime})^2}{
B}\right]
-r^2B^{\prime}\right\}_{\Sigma} \label{III4}.
\end{eqnarray}
>From (\ref{II7}), (\ref{II8}) and (\ref{III3}) we have
\begin{eqnarray}
\left\{\ddot{C}_2r^2+\ddot{C}_3-\frac{3}{2}\frac{(\dot{C}_2r^2+\dot{C}_3)^2}
{C_2r^2+C_3}-
\frac{\dot{C}_1r^2(\dot{C}_2r^2+\dot{C}_3)}{C_1r^2+1}
-2(\dot{C}_3C_1-\dot{C}_2)r\right. \nonumber \\
\left. +2\frac{(C_1r^2+1)}{C_2r^2+C_3}
\left[C_2(C_2-2C_1C_3)r^2+C_3(C_1C_3-2C_2)\right]\right\}_{\Sigma}=0.
\label{III5}
\end{eqnarray}

\section{A simple model}
A simple approximate solution for the functions $C_1(t)$, $C_2(t)$ and
$C_3(t)$ satisfying
the junction condition (\ref{III5}) is
\begin{equation}
C_1=\epsilon c_1(t), \;\;\; C_2=0, \;\;\; C_3=\frac{a}{t^2}, \label{IV1}
\end{equation}
where $0<\epsilon\ll 1$ and $a>0$ a constant. When $c_1=0$ then (\ref{IV1})
describes a collapsing Friedmann dust sphere, with $k=0$, whose radius
diminishes from arbitrarily large values until, at $t=0$, a singularity is
formed. The time $t$ runs from $-\infty$ to 0 and the constant $a$ is
proportional to the total mass inside the radius $r$. Substituting
(\ref{IV1}) into (\ref{III5}) we obtain up to $O(\epsilon)$,
\begin{equation}
\dot{c}_1+\left(\frac{t}{r^2_{\Sigma}}+\frac{2}{r_{\Sigma}}\right)c_1\approx
0,\label{IV2}
\end{equation}
which after integration yields,
\begin{equation}
c_1\approx
c_{1}(0)\exp\left(-\frac{t^2}{2r^2_{\Sigma}}-\frac{2t}{r_{\Sigma}}\right).
\label{IV3}
\end{equation}
Substituting the solution (\ref{IV1},\ref{IV3}) into (\ref{II6}-\ref{II8})
we obtain
\begin{eqnarray}
\kappa\mu\approx\frac{12a^2}{t^6}\left(1-\epsilon
2c_1r^2\right),\label{IV4}\\
\kappa p\approx\epsilon
\frac{4a^2c_1}{t^4}\left[1-\left(1+\frac{2r_{\Sigma}}{t}\right)\frac{r^2}
{r_{\Sigma}^2}\right],\label{IV5}\\
\kappa
q\approx-\epsilon\frac{8a^3c_1r}{t^7},\label{IV6}
\end{eqnarray}
which satisfy plausible physical conditions. It should be observed,
however, that in the general case $\epsilon \neq 0$, the range of $t$ is
restricted by physical considerations. Thus
for example if we want the the central pressure not to exceed the value of
the central energy density, then we should have,
\begin{equation}
\frac{3}{t^2}>\epsilon c_{1}.
\label{limit}
\end{equation}

 We see from (\ref{IV4}) that the energy density diminishes to
the outer regions due to
dissipation; from (\ref{IV5}) we have that pressure diminishes too towards
the outer regions while from (\ref{IV6}) we have that the heat flow
increases in that same direction.

The mass function (\ref{I11}) inside a radius $r$ with
(\ref{IV1}) and (\ref{IV3}) becomes,
\begin{equation}
m(r,t)\approx\frac{2r^2}{a}\left(1-\epsilon
2c_1r^2\right),\label{IV7}
\end{equation}
showing that dissipation diminishes the mass inside $r$. Now calculating
the rate of collapse (\ref{I5a}) with (\ref{IV1}) and (\ref{IV3}) we obtain
\begin{equation}
\Theta\approx\frac{6a}{t^3}\left(1-\epsilon c_1r^2\right),
\label{IV8}
\end{equation}
implying that dissipation slows down collapse. This result agrees with the
fact that $m(r,t)$ is diminished by dissipation.

The effective adiabatic index
\begin{equation}
\Gamma =\frac{d\ln p}{d\ln\mu},\label{IV9}
\end{equation}
gives a measure of the dynamical instability of the body at  given instant
of time. Calculating (\ref{IV9}) for $r=0$ and $r=r_{\Sigma}$ with
(\ref{IV1}) and (\ref{IV3}) up to the order
$O(\epsilon)$ in $p$ we obtain,
\begin{eqnarray}
\Gamma_{r=0}\approx\frac{2}{3}+\frac{t^2}{6r^2_{\Sigma}}+\frac{t}{3r_{\Sigma
}},\label{IV10}\\
\Gamma_{r=r_{\Sigma}}\approx\frac{5}{6}+\frac{t^2}{6r^2_{\Sigma}}+\frac{t}{3
r_{\Sigma}}.
\label{IV11}
\end{eqnarray}
We see from (\ref{IV10}) and (\ref{IV11}) that
$\Gamma_{r=0}<\Gamma_{r=r_{\Sigma}}$ which shows that the centre is more
unstable than the surface region of the collapsing body. This
 conclusion too agrees with our previous analysis.

\subsection{Calculation of the temperature}
Finally it is worth calculating the temperature  distribution, $T(r,t)$,
for our model,
through the Maxwell-Cattaneo heat transport equation \cite{12, 13,14, 8,
Triginer},
\begin{equation}
\tau
h^{\alpha\beta}w^{\gamma}q_{\beta;\gamma}+q^{\alpha}=-Kh^{\alpha\beta}
(T_{,\beta}+Ta_{\beta}), \label{V1}
\end{equation}
where $\tau$ is the relaxation time, $K$ the thermal conductivity and
$h^{\alpha\beta}=g^{\alpha\beta}+w^{\alpha}w^{\beta}$ the projector orthogonal
to $w^{\alpha}$.
Considering (\ref{I2}-\ref{I4}) then (\ref{V1}) becomes
\begin{equation}
\tau(qB)\dot{}B+qAB^2=-K(TA)^{\prime}. \label{V2}
\end{equation}

Substituting (\ref{II2}), (\ref{II4}) and (\ref{II8}) into (\ref{V2}) and
considering
$C_2=0$ we obtain, up to order $\epsilon$
\begin{eqnarray}
\tau (C_1C_3\dot{C}_3)\dot{} r+C_1\dot{C}_3r=-\frac{\kappa K}{4}
\left[T(C_1r^2+1)\right]^{\prime},
\label{1}
\end{eqnarray}
 Now, in the non--dissipative case (
$C_1=\epsilon c_1=0$) it follows at once
from (\ref{1}) that $T=T_{0}(t)$, implying that in that case the
temperature is homogeneous within the fluid distribution. Therefore, in the
general dissipative case $C_1 \neq 0$,
we shall have
\begin{equation}
T=T_{0}(t)+\epsilon T_{\epsilon}(r,t),
\label{temp}
\end{equation}
Then introducing (\ref{temp}) into (\ref{1}) we obtain up to $O(\epsilon)$
\begin{equation}
T\approx T_{c}+\epsilon c_{1}\left(
\frac{4a}{\kappa Kt^3}-T_{0}\right)r^2 -\epsilon\frac{4 a^2 \tau
c_{1}}{\kappa K t^5}
\left(\frac{t}{r^2_{\Sigma}}+\frac{2}{r_{\Sigma}}+\frac{5}{t}\right)r^2.
\label{VT3}
\end{equation}
where we have assumed for simplicity $K=$constant and $T_c(t)$ denotes the
central temperature.
The second term on the right hand side of expression (\ref{VT3})  exhibits
the influence of dissipation on the decreasing of temperature (remember
that $t<0$ ) with respect to the
non--dissipative case, as calculated from the non--causal (Landau--Eckart)
\cite{10,11}
transport equation, whereas the last term describes the contribution  of
relaxational effects. The relevance
of such effects have been brought out in recent works (see
\cite{Wagh,relax} and references therein). In particular it is worth
noticing the increasing of the spatial inhomogeneity of temperature
produced by the relaxational term, an effect
which has been established before \cite{HS97}.

\section{Conclusion}
We have presented the general field equations for a spherical dissipative
shear-free collapse. Some general properties concerning the effects of
dissipation on the
 collapsing body and its mass were discussed. By imposing conformal
flatness we showed that the system is completely soluble in its radial part
and producing three arbitrary time
functions. Then we matched this solution to the outgoing Vaidya radiating
spacetime. A simple model with a Friedmann limit is constructed satisfying
the junction conditions.

Besides
its simplicity, the merit of the model resides in the fact that it
exhibits in a very clear way the influence of relaxational effects on the
temperature, and thereby on the evolution
of the system.

It is also worth noticing the appearance of density inhomogeneities
directly related to dissipation, even though the space--time remains
conformally flat.
This reinforces doubts \cite{Bo} on the proposal that  the Weyl tensor
\cite{Pe} or some functions of it \cite{Wa}, could  provide a gravitational
arrow of time. The rationale
behind this idea being that tidal forces tend to make the gravitating
fluid more inhomogeneous as the evolution proceeds, thereby
indicating the sense of time.

\end{document}